\documentstyle[11pt,aaspp4,psfig,epsfig]{article}  % aaspp4.sty with figures
\begin{document}

\title{The core structure of AGN:\\ perils of adaptive optics artifacts}
%\title{The core structure of AGN:\\ A spiral starburst in NGC3227\\ An IR counterpart to a radio loop in NGC2992}
% line breaks possible with \\

\author{\sc 
Scott\ C.\ Chapman, Gordon\ A. H.\ Walker}
\affil{ University of British Columbia,
        Vancouver, B.C.~V6T 1Z4,~~Canada \\
{\it schapman@astro.ubc.ca,\ walker@astro.ubc.ca\/} 
}

% uncomment the above and below if two authors
\and   

 \author{\sc 
 Simon\ L.\ Morris}
 \affil{ Dominion Astrophysics Observatory, Victoria, B.C.,~~Canada \\
{\it simon@dao.nrc.ca\/ }}
\bigskip
\centerline{\today}

\bigskip
 
\begin{abstract}
As part of  a project
to map
nearby Seyfert galaxies in the near-IR
with adaptive optics,
we present high spatial resolution, near-IR images in J,H, and K of the
nuclei of NGC~3227 and NGC~2992, obtained with the Adaptive Optics Bonnette (AOB) on
CFHT. The $\sim0.15"$ resolution allows us to probe structures in the
core region at unprecedented scales. With NGC~3227, we are  able to identify
an inwards spiraling starburst in all three near-IR bands. NGC~2992 shows
evidence for emission along a radio loop.  
Compared with HST optical images, dust obscuration becomes significantly less pronounced at  longer wavelengths,
revealing the true geometry of the core regions.
The observed  structures may help to elucidate
fueling mechanisms for the central engine, as well as providing insight into
the unification paradigms.
The results are tempered with the discovery of AOB-related artifacts in the 
central arcsec of observed AGN galaxies which take on the appearance of
spiral/disk structures.
\end{abstract}

\section{INTRODUCTION}   % use capital letters for section names
\label{intro}            % \label{} and \ref{} is for cross-references
It is generally accepted that pronounced activity in
galaxies hosting Active Galactic Nuclei (AGN) results from accretion onto
a supermassive black hole.  This paradigm has led to a plethora of
research into AGN, of which the problems of overcoming the angular momentum
barrier to fuel the nucleus and
unification of the AGN types have risen to the forefront as especially
vexing and controversial. 
Near-IR  imaging 
has proven to be  a powerful means to  
study  these AGN problems since the dust  extinction   is  reduced,
and the contrast between the central AGN and the underlying stellar population
is improved.
A fresh perspective can be gained on the core regions of AGNs through the
high resolution
images now possible thanks to adaptive optics (AO).

We have begun  a project
to map 
nearby Seyfert galaxies in the near-IR
with adaptive optics
in order to study core morphologies and address possible fueling mechanisms.
Here, we present observations of  NGC~3227 and NGC~2992 obtained  with the AO
system  on the 3.6 m Canada-France-Hawaii
Telescope. Both galaxies live in disturbed environments:\\ 
$\bullet$ NGC~3227 is an SABa galaxy classified as Sy1.5, interacting with its dwarf elliptical neighbor,
NGC~3226. It has been much studied in recent years as it contains many of the
elements thought to be related to the formation and evolution of active
nuclei: emission line regions excited by both starburst and AGN continuum,
strong interaction, and a stellar bar (Gonzales Delgado \& Perez 1997, Arribas \& Mediavilla 1994).\\
$\bullet$ NGC~2992 is an Sa galaxy seen almost edge-on and interacting with NGC~2993.
It possesses an active Seyfert nucleus classified as Sy1.9.
A large and prominant dust lane runs through the center of the galaxy
roughly north to south,
splitting the nuclear region in two.
Ulvestad and Wilson (1984) found that the radio structure of the nucleus of
NGC~2992 has the shape of a ``figure-8'', with a maximum extent of about
2000~pc, oriented out of the plane of the galactic disk.
Most of of the 6cm radio emission from the center of the galaxy arises in the
loops of the figure-8 rather than in the nucleus.

\section{OBSERVATIONS}          
\label{sec2}

The imaging observations were obtained at the CFHT in March, 1997, using the MONICA
near-IR camera (Nadeau et al. 1994) mounted at the f/20 focus of the
Adaptive Optics Bonnette (AOB). The
detector is a Rockwell NICMOS3 array with 256x256 pixels and a 0.034"/pix
scale.
The CFHT AOB is based on curvature wavefront sensing (Roddier 1991),
and uses a  19 zone bimorph
mirror to correct the wavefront distortions.
As the field size is small (9"x9"), blank sky images were taken intermitantly
between science frames. On-source images were taken in a mosaic of 4 positions,
alternately putting the galaxy core in each of the four quadrants of the
array.
Flux and PSF calibrations were performed using the UKIRT standard stars
fs13 and fs25. Flat-field images were taken on the dome with the lamps turned
on and off to account for the thermal glow of the telescope. The nuclei
of the galaxies themselves were used as the guiding source for the AO system,
a clear point source for NGC~3227 and a more extended core for NGC~2992. The natural seeing averaged 0.6"-0.8"
throughout the observations resulting in relatively high strehl ratios in
all bands, and FWHM of 0.14", 0.17", 0.22" at K, H, J bands respectively.
At a distance of
15\,Mpc for NGC\,3227, 1"=76pc using $H_0=50$.
The estimated distance of NGC\,2992 at 20Mpc implies an angular scale of
100pc / 1".

\section{THE COMPLETE SAMPLE}
\label{sec3}
Candidate galaxies for our survey were chosen based primarily on core guidability with AOB.
Our sample galaxies therefore tend
to have the brightest cores amongst nearby Seyferts (though not necessarily unresolved)
and may not be representative of the lower luminosity active nuclei.
However, our Seyferts typically have bright X-ray flux
and are perhaps the best sample with which to study
 the AGN phenomenon, since they are uncontaminated with borderline AGN galaxies.
There are roughly equal numbers of Seyfert type 1 and 2 galaxies, since a 
bright extended core seems to provide a good guiding source with  
curvature sensing adaptive optics.
We note however that the types of artifacts around the central AGN spike
vary with the core morphology, and identifying physical  structures is a 
nontrivial matter within the central arcsecond.
The complete sample is presented in Table 1, along with alternate designations,
redshift, filters obtained, Seyfert class (S=saturated core, U=unresolved core,
R=resolved core; all Sy2 are resolved; (S) = HST saturated core); galaxy
classification, and camera used: KIR or MONICA.

\subsection{AOB artifacts - the difficulty of identifying physical structure}

The exciting results on AGN cores emerging from AOB studies are tempered with the discovery of achromatic, AOB-related artifacts in the
central arcsec surrounding the bright nuclei of observed AGN galaxies.
These artifacts masquerade convincingly 
as the expected types of morphologies in the centers of Seyfert galaxies: 
small-scale spiral arms, edge-on disks or tori, double nuclei, and outflows.
Figure 1 shows a representative sample of AGN cores.
Elliptical isophotes were fit to the galaxy core region. A smooth model was then
built from the resulting fit, and subtracted from the raw image. 
Alternatively, median filtered images were subtracted from the core with the
same result. 
The most pronounced effect in all the raw images is an extended core to the
North.
Figure 1 clearly depicts that these structures exist at the same scale with
similar shape and orientation relative to North, regardless of
Seyfert type or distance to the galaxy. In particular NGC\,3998, NGC\,3516,
and Mkn\,348 all show a cross-like pattern suggestive of small-scale
spiral arms plus a bar or edge-on torus. 
The K-band image of NGC\,3998 is
displayed in Figure \ref{artfig2}, and shows the effect of this cross pattern artifact is
significant even in the raw image. 
The northern component of the cross encompasses more than 10\% of the peak of the central core.
In the case of NGC\,3227, the AOB core was compared with HST-NICMOS data 
(Alonso-Herrero, private communication) and the core cross pattern was found
not to exist, even at very low signal levels. 
Note that a similar feature 
may have been  mistakenly identified as a physical structure in NGC1068 (\cite{rouan98}).
We stress, however,   
that the AOB system is perfectly reliable in identifying systems such as 
binary stars or galaxies with true double nuclei, if the relative flux is 
less than 
a factor of 10, or if the artifact is carefully taken into account 
(see \cite{knap97} and \cite{lai98} for  resolved double nuclei in
Mkn273 and Mkn231 respectively which appear to be real). 

A technique has been developed to reconstruct the AOB PSF from the modal control used during the actual observations (\cite{veran98}). The algorithm requires
a fairly bright guiding source to accurately reproduce the PSF, and not 
all our galaxy cores satisfy this criterion.   
Reconstructed PSFs for  4 of the galaxies depicted in Figure \ref{artfig1} 
(all with bright cores), are shown in figure \ref{artfig3} with a smooth 
model subtracted.
Structures are similar in scale to the actual galaxy data, showing some kind
of ``lobe" configuration. The actual galaxy counterparts to these reconstructions display a broad range in core morphology from the extremely bright pointsource in NGC\,4151 to the bright but extended Seyfert 2 Mkn\,620.  The reconstructed PSF has essentially
the same morphology in all cases, and certainly does not reproduce the
actual artifact morphology in the galaxy cores.
However, the reconstruction does not take into account the extended nature of
the underlying galaxy and the nuclear core, and would not be expected to
reproduce the artifact if it was related to this. Non-common light path to the
wavefront sensor and IR array may also contribute to the discrepancy.
%But does it reproduce the stellar artifact? NO.

The structures are most prominent
when a bright point source dominates the core, and is less apparent in the
weaker or more extended cores, possibly indicating that the artifact is
strongest when the AOB correction is at its best. The morphology is also not
dependent on the camera used with AOB (both the 1k$^2$ pixel KIR and 256$^2$ pixel
MONICA show identical artifacts, once the images have been rotated to the
same orientation on the sky).
 
Although some stars imaged with AOB were reported not to show extended
PSF cross-patterns, we find that the structures
appear to contaminate our calibration stellar images corrected with AOB in a 
similar manner
(Chapman et al. 1999a in preparation). 
Thus the extended underlying galaxy is likely 
not the primary cause of the cross-shaped and double/triple-lobed artifacts. Further 
modeling of the response of the curvature
based wavefront sensing AO system is required to ascertain the root of
such structures.
Deconvolution techniques
may be able to account for these artifacts sufficiently, once their true nature is uncovered. 
This will allow the central arcsecond
of active galaxies to be probed with more confidence using AOB.

\begin{table}
\begin{center}
\caption{ \hfil The AOB observed Seyfert galaxy sample \hfil }
\begin{tabular}{lcccccccl}
\noalign{\medskip}
Galaxy & Alternate & z & filters & sy class & gal type &
instrument \cr
\hline
\noalign{\medskip}
ic 4329a & eso 445-g50 & 0.016 & J,H,K,CO & S1 & Sa,S0 & KIR \\
mrk 1066 & UGC 2456 & 0.012 & J,H,K & 2 & Sc,SB0 & KIR \\
mrk 1330 & NGC 4593 & 0.009  & J,H,K & U?1 & Sb/c, SBb & KIR \\
mrk 3  & UGC 3426 & 0.014  & J,H,K & 2 & S0 & KIR \\
mrk 620  & NGC 2273 & 0.006  & J,H,K & 2 & SBb,SBa & KIR \\
mrk 744  & NGC 3786 & 0.010  & J,H,K & U1.8(S) & Sb,Sa &KIR \\
mrk 766  & NGC 4253 & 0.012  & J,H,K & U1.5(S) & SBc,Sa &KIR \\
ngc 2639  & NGC 4593 & 0.011  & J,H,K & R1 & Sb,Sa & KIR \\
ngc 2992  & & 0.007  & J,H,K,CO & 2 & Sa & KIR/MON \\
ngc 3516  & UGC 6153 & 0.009  & J,H,K & S1.5 & S0 & KIR \\
ngc 3998 &  & 0.008   & J,H,K & U1.5 & E  & KIR \\
ngc 4051  & & 0.002  & J,H,K & S1 & Sb,SBbc &KIR \\
ngc 4151  &  & 0.002  & J,H,K & S1 & Sa/ & KIR \\
ngc 4968  & ESO 508-g6  & 0.009  & H & 2 & Sa/SB0a &KIR \\
ngc 5033 &  & 0.002  & J,H,K & 2  &  & KIR/MON   \cr
ngc 5135 & ESO 444-g32 & 0.013  & K & 2 & Sc,SBaba & KIR \\
ngc 5273 & & 0.004  & H & R1.8 & Sa/  & KIR   \cr
ngc 3081 & & 0.007  & K & 2 & SBca & KIR  \cr
ngc 3393 & & 0.012  & K & 2 & Sa/SBa & KIR \\
mrk 1376 & NGC 5506 & 0.007  & J & R1.9 & Sa/edge & KIR\\
ngc 3227 & UGC 5620 & 0.003   & J,H,K & U1.5(S) & SBa & MONICA \\
ngc 5548 & MRK 1509 &0.017  & J,H,K & U1.5(S) & Sa & MONICA \\
mrk 348 & NGC 262 & 0.014  & H,K & 2 & Sa,S0a & MONICA \\
ngc 1068 & & 0.003  & H & 2 & Sb/ & MONICA   \cr
ngc 7469 & & 0.016  & H & U1(S) & Sb/c & MONICA  \cr
ngc 1241 & & 0.013  & H & 2 & Sb/c, SBb & MONICA  \cr
ngc 1275  & & 0.013   & K & 1.5 & Sa/ & MONICA   \cr
ngc 1386  & & 0.002  & H,K & 2 & Sb/c & MONICA \\
ngc 5728  &  & 0.005 & K & 2 & Sb/ & MONICA   \cr
ngc 5929  & UGC 9851 & 0.008  & J,H,K & 2 & S0,Sab & MONICA  \cr
ngc 5953  &  & 0.007  & H & 2 & Sb/ & MONICA   \cr
ngc 6814  &  & 0.003   & J,H,K,CO & U1  & Sa/ & MONICA   \cr
ngc 7465  & MRK 313 & 0.006  & J,H,K & 2 & Sa/ & MONICA   \cr
ngc 7582  & ESO 291-g16 & 0.005  & H,K & 2 & Sa/ & MONICA   \cr
ngc 7590 &  & 0.005  & H & 2 & Sb/ & MONICA   \cr
ngc 7743  &  & 0.007  & H,K,CO & 2 & Sa/S0a & MONICA   \cr
ngc 5005  &  & 0.002  & H & 2  & S0/ & MONICA   \cr
\end{tabular}
\end{center}
\end{table}

\section{NGC\,3227}
The CFHT  J,H,K- and HST V-band images are presented in Figure \ref{n3fig1} on a log scale.
A diffuse, ellongated structure containing wispy spiral bands is seen surrounding the nucleus in all wavelengths.
Subtraction of a smooth model reveals that this region is punctuated with bright
 knotty structures tracing out a mini-spiral pattern within a region 3"x2" 
(Figure \ref{n3fig3}). 
The colours of these knots are consistent with a red
supergiant population, with a scattered AGN light contribution near the core.
However, the presence of the AOB artifacts in the very core region make the
identification of physical knots difficult.
We explored several methods of removing the low frequency galactic component, including various smoothing filters, a one-dimensional elliptical isophote model, and a multi-component (bulge+disk+point source) elliptical isophote model.
All methods consistently unveil the knotty  spiral structure.
However, in the central arcsec of the galaxy,
subtracting 
isophotal fitting models results in prominent artifacts which obscure structural
details as described in section 3.1.

Color maps are formed by convolving the images to the worse resolution of 
a given pair and taking the flux ratio (Figure \ref{n3fig2}).
Any color gradients in these images  can result from several different
processes: 1) change in dust 2) change in stellar population 3) change in gas.
The most prominent feature is an irregular-shaped patch to the southwest.  The fact that this region appears clearly as a deficit in the V-band image, and takes on a 
patchy morphology is strong evidence for dust obscuration as the source of the color gradient. The  region is therefore  most pronounced the V-K color map, since the
K image is least affected by dust. The J-K image
indicates that substantial dust still affects this region in the J-band.
The nucleus is also very red relative to V and J, possibly as a result of thermal dust emission
in the K band.
The red colors of the knotty spiral starburst stand out from a  region
slightly bluer than the larger scale bulge of the galaxy. 

The images are distorted by PSF artifacts in the central 0.5", with strong diffraction spikes in the HST image.
However, there appears to be 
a disk-like feature most clearly visible
in the J-K and V-K color maps, as it seems to be bluer than the rest of the galaxy at K. The position angle of 43$^\circ$ indicates that this elongation is
unrelated to the AOB artifacts seen in figure 1.
The 1D profiles of the galaxies are similar at J, H and K,  displaying bumps
in ellipticity at 1.5 and 0.5 arcsec radius, confirming the presence and position
angle of the above disk and the enhanced region coincident with the spiral
starburst. The isophotes are twisted of order 10$^\circ$ in both cases.

The images are also compared to the 6cm and 18cm MERLIN radio continuum emission, both of which align with the axis of the nuclear spiral as seen in Figure \ref{n3fig3}.
Previous explanations for the
radio structure (Mundell et al 1995), invoked the standard unified AGN model
to explain this emission as collimated outflow. However, there is an 
offset in orientation of the [OIII] ``cone"
and the small-scale radio features.
A projection effect would be possible, but this would necessitate that the
NE side of the disc is closer to us than the SW side. This could only occur 
if the spiral arms were leading rather than trailing (Mundell et al 1995).
As figure \ref{n3fig3} shows, the radio aligns well with both the orientation and some
of the knots of the starburst spiral, and the obvious interpretation is that
we are seeing synchrotron emission from SNe remnants.

\subsection{Discussion}

\begin{table*}
{\scriptsize
\begin{center}
\centerline{Table 2}
\vspace{0.1cm}
\centerline{The structural components of NGC\,3227}
\vspace{0.1cm}
\begin{tabular}{llllll}
\hline\hline
\noalign{\smallskip}
{Component} & {Scale} & {PA ($^\circ$)} & {Ellipticity} & 
{Observed with} & {Function in galaxy} \cr 
\hline
\noalign{\smallskip}
Large Scale galaxy & 1-10 kpc & -25 & 0.5 & V-band & \cr
Large Scale bar & 1-5 kpc & -45 & 0.8 & galaxy subtracted & funnel material to ILR at 7"\cr
Extended [OIII] & 1kpc & 35 &  & OIII filter/ OASIS [SIII] & collimated emission? \cr
Circum-nuclear ring & 1 kpc &  &  & H$_\alpha$ & ILR\cr
Medium Scale bar & 100-1000pc & ? &  & submm CO  & funnel material to ILR at 2"\cr
\noalign{\smallskip}
K-band ellipse & 200 pc & -10 & 0.2 & model subtract/ color maps&
spiral starburst, ILR\cr
Radio jets/blob & 100 pc & -10 &  & MERLIN 8/16cm& SN/outflow?\cr
K-V blob & 100 pc & -10 & 0.1 & K-V map, raw Vband & bluer than galaxy \cr
K-J annulus & 100 pc & 40 & 0.3 & K-J, K-V, J maps & twisted disk/bars, scattered AGN
 light \cr
\noalign{\smallskip}
\noalign{\hrule}
\noalign{\smallskip}
\end{tabular}
\end{center}
}
\end{table*}

Several possible scenerios emerge from these results. We tabulate the observed
structures in this galaxy from the largest to the smallest in Table 2.
On the largest scales
Gonzalez Delgado et al. (1997) noted that a large-scale bar appears to transport
 material
towards an inner radius which corresponds to the calculated inner Linblad resonance (ILR) at
roughly 7".
At this point, prominent dust and HII regions indicate
substantial star formation.
Within this region, a molecular bar of length $\sim1$kpc is observed in CO with
and ILR of 2" (Schinnerer 1998). This radius corresponds with the outer extent
of the spiral starburst rings in our images.

With such nested bar structure repeating itself at these two larger scales,
it is natural to speculate that the small scale elongation seen in the color
maps and profile analysis may be yet another bar potential funneling material
down to the scales where viscous forces may take over to fuel the AGN.
However, a larger scale extended [OIII] region (Arribas et al. 1995) 
lies to the northeast and has been
interpretted as a narrow-line region ionized by the AGN, collimated into
a bi-cone by a small-scale ($\sim pc$) dusty torus. The fact that this
collimation axis roughly aligns with our ``bar" may be an indication that
this elongation actually represents scattered AGN light. This is made all
the more convincing by the blue colour of the elongation.

On the other hand, if our observed small-scale elongation is some sort of twisted disk as found in Centaurus\,A by Schreier et al. (1998), its plane lies roughly perpendicular
to the axis defined by the radio ``jet" observed at 6 and 18\,cm, and would
be consistent with a collimated radio jet normal to an accretion disk plane.
For the radio emission to be interpretted as an outflow, 
the collimated [OIII] ionization
picture would then have to be abandoned. 
A more detailed analysis can be found in Chapman et al. (1999b).

\section{NGC\,2992}

In figure \ref{n2fig}, we present an H-band AOB image of the central 7" of
NGC\,2992. Of obvious note are the jet-like extent to the NW, and the elongated
isophotes to the SW along the galaxy disk.
The V-band HST image is also depicted with 8.4GHz radio contours overlaid. 
Although it is still clear that the isophotes are elongated to the SW along
the galaxy disk, the galaxy morphology is much more distorted due to the
effects of dust and there is no indication of the extension to the NW out
of the galaxy plane. 

The V-band image shows that the radio emission lying along the galaxy disk
has no obvious opical emission associated and appears to lie well within
the dust lane.
When we form an H-V color map, we now observe highly reddened emission lying
along the radio contours in the disk of the galaxy, however with what appears to be
a loop to the north, associated with an inner loop of radio emission.

By subtracting a model image consisting of either elliptical fitted isophotes,
or a smooth median filtered image, we are able to discern the spiral arms along the
disk, as well as an extension  to the West (figure \ref{n2fig}), 
also noted 
in \cite{alonso98}. There is clearly some radio emission coincident with the
southern spiral arm, which breaks up into a similar knotty morphology to
the H-band model-subtracted image. 
At faint
levels, it is also possible to discern knotty features along the northern red H-V loop, the
colors of which are consistent with star formation. 
The extended feature aligns with the beginning of the
H-V loop noted above, but continues outward to fill the radio loop. 
%that the red loop is in fact much weaker H-band emission which  

There is,  however, little sign of optical or near-IR counterparts to the
radio loops out past the disk of the galaxy, even at K-band  where the 
ability to see through the dust lane is greatest.
This is evidence for the actual figure-8 loops lying out of the galactic disk
plane, superimposed over fairly strong disk emission related to 
star formation. If the extended near-IR emission to the West is related to 
the NW radio loop, the dust lane must extend out past the loop in order
to heavily obscure the optical/near-IR emission.
This sort of near-IR ``jet" may exist towards the SE radio loop as well, 
although
largely obscured by the galaxy disk.

%However this does not strongly constrain the angle of emission 
%We thus conclude that the BIG radio loops are an entity unto themselves,
%since there is no evidence even at high spatial resolution in K-band,
%for any such related emission.

There are several favored models for such figure-8 radio emission.
The most convincing in light of our new near-IR imaging is that the 
loops result from some sort of expanding  gas bubbles
which are seen preferentially as limb-brightened loops (Wehrle 1988).
Such outflows may be associated with the AGN core, which is consistent with
the orientation of the proposed ionization cones observed at larger scales 
Allen et al. 1998.
Superwinds in starbursts  would blow preferentially out of the galaxy plane, such as
in NGC253 (Unger et al. 1987), providing a similar mechanism even without a
strong AGN driving the outflow. Here, however, the [OIII] emission is likely to be associated, but not continuum emission
thus our data likely rule out this latter model.
These possibilities are further explored in Chapman et al. (1998c)

%The K-band is more round.
%This can be seen successively in the color maps as there is a bright line
%going out from J-V to K-V, moving to the West.

%With the figure-8, if we take out the "line" along the V-band dust lane, then
%the figure-8 is still there and maybe even a *better* figure 8.
%There does appear to be excess near-IR emission along the dust lane, 
%traced by the NE-SW arms of the radio figure-8.
%In addition there is emission which loops around through the dust lane to the north, and with knots along the inner radio loops to the north and south.

%Q: Are the colors consistent with rapid SB with associated SN (as with NGC3227)?
% Thus are the NE-SW line of radio emission distinctly different from the NW-SE
%radio emission, and does this then imply that the *inner loops* identify with
%the BIG radio loops, out of the galaxy plane.

%near-IR jet? seen extending to fairly large scales, filling the region inside
%the northern radio loop. This may exist to the SE radio loop as well, although
%largely obscured by the galaxy disk. (note the reddening to the E and W from X
% - more to the E?).

%Pedlar et al. Blowing up bubbles using "outflows" - if this is AGN-related,
%then this could be collimated due to torus, even if AGN now dead in this
%galaxy.

%
% FIGURE 1 : mosaic of artifacts
%
\begin{figure}
\caption{A representative sample of the cores of our Seyfert galaxies at
 K-band, 3.4" on a side. An elliptical isophote 
model has been subtracted from each.
Left to right: Ic4329, Mkn1330, Mkn744, Mkn620, NGC3516, NGC3998, NGC4051,
NGC4151, Mkn348}
\label{artfig1}
\end{figure}

\begin{figure}
\caption{NGC3998 K-band, 3.4" on a side, raw data. A cross-type feature
surrounds the nucleus, resembling spiral and torus structures. }
\label{artfig2}
\end{figure}

\begin{figure}
\caption{Four of the reconstructed PSFs from Figure 1 with a smooth model subtracted (4" on a side). 
Structures are similar in scale to the actual galaxy data, showing some kind
of ``lobe" configuration. Left to right: Mkn620, NGC3998, NGC4051, NGC4151}
\label{artfig3}
\end{figure}

\begin{figure}
\caption{Clockwise:
 ngc\,3227 HST F606W "V-band", K, H, J-band images. The AOB images
have been deconvolved using the LUCY algorithm with 20 iterations.}
\label{n3fig1}
\end{figure}

\begin{figure}
\caption{ngc\,3227 color maps, clockwise: K-V, J-V, H-V, K-J}
\label{n3fig2}
\end{figure}

\begin{figure}                                    % fig.1
\caption{ngc\,3227 model subtracted images (Top: V-band, Bottom: K-band 4" on
a side),  with MERLIN radio contours to right (18cm) and left (6cm). The
18cm peaks appear to align with the galaxy core and a starformation knot to the north.}
\label{n3fig3}
\end{figure}

\begin{figure}
\caption{a) NGC\,2992 H-band, 7" on a side;
b) V-band HST image with 8.4GHz radio contours overlay;
c) H-band smooth model subtracted;
d) H-V color map: red is bright at H-band.}
\label{n2fig}
\end{figure}


\begin{thebibliography}{dum}
% References are quite complex and a bit unplasant, as you must
% supply the code for each paper, something like below. Use the
% journal codes, as in aastex.tex
\bibitem[Allen et al. 1998]{allen98} Allen et al. 1998, astro/ph9809123
\bibitem[Alonso-Herrero A. et al. 1998]{alonso98} Alonso-Herrero A., Simpson C., Ward M.\,J., Wilson A.\,S., 1998, ApJ, 495, 196
\bibitem[ant93]{a93} Antonucci R., 1993, \aapr 31, 473
\bibitem[]{} Chapman S. \& Morris S.L., in preparation 1999a.
\bibitem[]{} Chapman S. \& Morris S.L., 1999b, submitted to MNRAS.
\bibitem[]{} Chapman S.,  Morris S.L., Alonso-Herrero A., Falcke H., in preparation 1999c.
\bibitem[Falcke et al.]{falcke98} Falcke, H. et al., 1988, AJ, 95, 1689
\bibitem[ant93]{ar94} Arribas S., Mediavilla E. 1994, \apj, 437, 149 
\bibitem[ant93]{gonz97} Gonzalez Delgado R.M., Perez E. 1997, \mnras, 284, 931 
\bibitem[Knapen et al. 1997]{knap97} Knapen J., et al, 1997 \mnras, 500, 500
\bibitem[Lai et al. 1998]{lai98} Lai O. et al., 1998, A\&A, in press.
\bibitem[ant93]{mun95} Mundell C.G., et al. 1995, \mnras,  275, 67
\bibitem[ant93]{ruc97a} Nadeau D., Murphy D.C., Doyon R., Rowlands N., 1994, \pasp 106, 909
\bibitem[ant93]{ruc97a} Roddier F.J., Graves J.E., McKenna D.  , Northcott M.J.  , 1991,
\procspie Vol.~1524, p.~248
\bibitem[Rouan et al. 1998]{rouan98} Rouan D. et al., 1998, A\&A, in press.
\bibitem[Schreier et al. 1998]{sre98} Schreier E.J. et al. 1998, \apjl, in press
\bibitem[Schinnerer et al. 1998]{se98} Schinnerer E. et al. 1998, AGM 14, H08
\bibitem[Ulvstad \& Wilson 1984]{ulv84sre98} Ulvstad \& Wilson A. 1984, 
\bibitem[Unger et al. 1987]{pedlar80} UNGER, S. W., PEDLAR, A.,
 AXON, D. J., WHITTLE, M.,
 MEURS, E. J. A., WARD, M. J., 1987 MNRAS, 228, 671
\bibitem[Veran et al. 1998]{veran98} Veran J.P., Rigaut F., Ma\^{\i}tre H., Rouan D., 1997,
Journ. of the Optic. Soc. of America A 14, 11
\bibitem[]{} Wehrle A.\,E., Morris M., 1988, AJ, 95, 1689
\end{thebibliography}
\end{document}